\documentclass{article}

\usepackage{amsmath}
\usepackage{graphicx}

\author{Aron C. Wall\footnote{aronwall@umd.edu}
\\ \textit{Maryland Center for Fundamental Physics} \\ \textit{Department of Physics} \\ \textit{University of Maryland} \\ \textit{College Park, MD 20740-4111, USA} }
\title{Proving the Achronal Averaged Null Energy Condition from the Generalized Second Law}
\date{\today}

\begin{document}

\maketitle

\begin{abstract}
A null line is a complete achronal null geodesic.  It is proven that for any quantum fields minimally coupled to semiclassical Einstein gravity, the averaged null energy condition (ANEC) on null lines is a consequence of the generalized second law of thermodynamics for causal horizons.  Auxiliary assumptions include CPT and the existence of a suitable renormalization scheme for the generalized entropy.

Although the ANEC can be violated on general geodesics in curved spacetimes, as long  as the ANEC holds on null lines there exist theorems showing that semiclassical gravity should satisfy positivity of energy, topological censorship, and should not admit closed timelike curves.  It is pointed out that these theorems fail once the linearized graviton field is quantized, because then the renormalized shear squared term in the Raychaudhuri equation can be negative.  A ``shear-inclusive'' generalization of the ANEC is proposed to remedy this, and is proven under an additional assumption about perturbations to horizons in classical general relativity.
\end{abstract}

\newpage

\section{Introduction}\label{intro}

The generalized second law (GSL) of black hole thermodynamics states that a certain sum of the area of a black hole's event horizon and the entropy outside of it cannot decrease over time \cite{hawking75}:
\begin{equation}
\frac{d}{dt}(\frac{A}{4G\hbar} + S_{out}) \ge 0.
\end{equation}
This law is an analogue of the ordinary second law of thermodynamics.  However, there is a crucial difference between the two.  The ordinary second law holds trivially for purely information theoretical reasons in any quantum field theory (QFT), whereas the generalized second law seems to also require some restriction on the stress-energy tensor of the quantum fields.  For if it were possible to send a sustained flux of negative energy into a black hole, the area of the black hole would shrink, without necessarily being accompanied by any corresponding increase in the outside entropy $S_{out}$, violating the GSL.  This threat to the GSL is made more credible by the fact that there are many ways to create local negative energy fluxes in QFT \cite{BM69}.  In fact, a black hole could not evaporate in the first place without some negative energy flux.  Thus, the truth of the GSL depends on the existence of some sort of constraint on the stress-energy tensor.  One such proposed constraint is the averaged null energy condition (ANEC) \cite{borde87}\cite{roman88}.

The ANEC states that along a given infinite null geodesic $N$, the integral of the null component of the stress energy is nonnegative:
\begin{equation}\label{anec}
\int^{\infty}_{-\infty} T_{ab} k^a k^b d\lambda \ge 0,
\end{equation}
where $T_{ab}$ is the expectation value of the renormalized stress-energy tensor, $\lambda$ is an affine parameter along the null geodesic which goes from $-\infty$ to $\infty$, and $k^a$ is a vector pointing along the direction of the null geodesic and satisfying $\lambda_{,a} k^a = 1$.  Matter violating the ANEC might be used to create traversable wormholes, which could in turn produce closed timelike curves and hence causality problems \cite{MTY88}\cite{FSW93}.  In Minkowski space, the ANEC has been proven for free scalar fields in any dimension, \cite{klink91}, free electromagnetic fields in four dimensions \cite{folacci92}, and for any QFT with a mass gap in two-dimensions \cite{verch00}.  The ANEC is also closely related to ``quantum inequalities'', which are bounds on the severity and duration of negative energy fluxes which have been proven for several specific quantum field theories \cite{FR95}.

However, exceptions to the ANEC do occur on curved spacetime backgrounds.  For example, if one dimension of space is compactified into a circle, the Casimir effect can produce a negative energy density, in which case a null geodesic which goes round and round the compactified dimension violates the ANEC \cite{klink91}.  Counterexamples can also be found for null geodesics on a Schwarzschild background in some field states \cite{visser96}.  What these examples have in common is that the null geodesic $N$ is chronal (i.e. it contains points which are connected by a timelike curve).  So it is still possible that the ANEC holds semiclassically on complete achronal null curves (known as ``null lines''\footnote{An null line is always a geodesic.  This can be seen by choosing locally inertial coordinates near any point $X$ at which the curve fails to obey the geodesic equation.  In these coordinates, the light ray's 3-velocity is changing with time, meaning that it cannot outrun all possible timelike curves in the neighborhood of $X$.}).  This version of ANEC has been proven by Wald and Yurtsever \cite{WY91} in the case of minimally coupled free scalar fields, either on a curved two-dimensional spacetime, or on a four-dimensional spacetime with a bifurcate Killing horizon.

Graham and Olum \cite{GO07} have proposed that the ANEC is true for all null lines on semiclassically ``self-consistent'' spacetimes, meaning that the semiclassical Einstein equation is satisfied.  They show that this weaker version of the ANEC is sufficient to prove both a topological censorship theorem that rules out traversable wormholes, as well as a couple of theorems that rule out the creation of closed timelike curves (cf. Ref. \cite{minguzzi08}).  It is also sufficient to prove a positive energy theorem for general relativity \cite{PSW93}.  These theorems provide some evidence that the achronal semiclassical ANEC is in certain respects a substitute for the (quantum-mechanically violated) null energy condition.  However, all of these theorems rely on a focusing theorem by Borde \cite{borde87}, which in turn depends on the assumption that the shear $\sigma_{ab}$ of any null congruence satisfies
\begin{equation}\label{shear}
\sigma_{ab}\sigma^{ab} \ge 0.
\end{equation}
Although Eq. (\ref{shear}) is trivial at the classical level, once fluctuations in the metric are quantized, $\sigma_{ab}$ has to be promoted to an operator.  $\sigma_{ab}\sigma^{ab}$ then becomes divergent, and requires the subtraction of an infinite quantity to be well-defined.  The resulting finite term can be negative, e.g. in the case of a black hole which Hawking radiates gravitons \cite{CS77}.  So a proof of the results in Ref's \cite{GO07} and \cite{PSW93} would seem to require a generalization of the ANEC which also places bounds on the degree to which Eq. (\ref{shear}) can be violated.  A possibly sufficient generalization will be argued for in section \ref{graviton}.

The goal of this article is to show that if the GSL holds, not only for global black hole horizons but on general ``causal horizons'', then for any null line on a curved background spacetime satisfying certain appropriate properties, any quantum fields on this background must also satisfy the ANEC on that null line.  More precisely: Consider a manifold $M$ equipped with a Lorentzian metric $g$ satisfying the Einstein equation, the null energy condition, and the ``slightly weaker extrastrong'' \cite{penrose99} version of cosmic censorship.  Given a null line $N$ on $M$, Galloway's ``null splitting theorem'' \cite{galloway00} shows that the null line lies on both a past and a future causal horizon $H$.  Now introduce quantum fields on $M$ whose stress-energy causes a small gravitational perturbation to the expectation value of the metric $g$ via the Einstein equation.  Assuming that the horizons persist under this perturbation,\footnote{Cf. section \ref{pert} for a comment on how the proof might be extended to cases in which the perturbation removes the horizon altogether.} there will then exist a perturbed past horizon $H_{past}$ and future horizon $H_{fut}$.  Then, provided there exists a suitable renormalization procedure to  define a finite generalized entropy due to the quantum fields, if $H_{fut}$ obeys the GSL, and $H_{past}$ obeys the time-reverse of the GSL, then it follows that the quantum fields on $M$ must satisfy the ANEC over $N$.  Note that no assumption has been made about either the spins of the quantum (non-metric) fields or their interactions with one another.\footnote{However, the argument as presented here does not apply to the case of fields which are non-minimally coupled to the metric, because in that case there are usually extra terms modifying the Einstein equation, the GSL \cite{wald93}, and possibly the ANEC as well.  In cases where one could remove the non-minimal coupling by means of a field redefinition, the argument would then apply to the new metric and stress-energy tensor.  Extension of this result to the case of nonminimal coupling would therefore be desirable.}

The ANEC result proven here is similar in scope to the ``self-consistent achronal ANEC'' conjectured by Graham and Olum \cite{GO07}.  The present result is weaker insofar as it is restricted to the case in which the gravitational perturbation to the background metric is small, but it is stronger insofar as it only requires the existence of a null line on the background spacetime, not on the perturbed spacetime.  This is a significant extension, because Graham and Olum have shown that there are no generic spacetimes which satisfy their condition except trivially (by having no null lines at all).

To control the size of the gravitational perturbation, I will use an expansion in Planck's constant $\hbar$.  The content of the ANEC can be defined using only QFT on the fixed classical background, while the content of the GSL will be defined using only the first order gravitational correction to the background metric and the non-gravitationally corrected quantum field entropy.\footnote{This approach differs from the perturbation expansion used by Flanagan and Wald \cite{FM96}, who expand in the difference between the quantum state and the vacuum rather than in $\hbar$, and assume that all the curvature of the spacetime comes from the quantum matter fields.}

The plan of this paper is as follows: Section \ref{gsls} describes the meaning of causal horizons and the GSL.  Section \ref{class} discusses the classical background metric $M$ and uses the null splitting theorem to place important restrictions on the spacetime.  Section \ref{pert} provides the details of the perturbation expansion of the metric in $\hbar$ required for the proof.  Section \ref{proof} proves the ANEC at leading order ($\hbar^{1}$) from the GSL together with the other assumptions listed above, but does not address the renormalization of the generalized entropy needed to make the proof rigorous, nor does it address situations in which gravitational fluctuations are important.  Section \ref{ren} discusses what properties a renormalization procedure must have in order for the proof to be valid, and argues that it is likely, though not certain, that such renormalization procedures exist.  Section \ref{graviton} considers the case in which gravitational fluctuations are non-negligible.  In this case there is a plausible generalization of the ANEC which includes an additional shear-squared term.  This proof is then extended to this ``shear-inlcusive'' ANEC if an additional assumption is made about the way past and future horizons respond to gravity waves in classical general relativity.  Under the additional assumption that the GSL applies regardless of the number of particle species in nature, the usual ANEC is also proven for null lines in this case.

\section{The GSL and its Time Reverse}\label{gsls}

The GSL appears likely to hold for a wide variety of horizons, including not only black hole horizons but also de Sitter and Rindler horizons.  Although the GSL is usually phrased as a statement about global horizons, there is good evidence for a more general formulation \cite{JP03}, suggesting that one should try to formulate the GSL for as general a definition of ``horizon'' as possible.  And in order to prove the ANEC for all null lines, I need a version of the GSL which can be applied to any spacetime containing a null line.

Following Jacobson and Parentani \cite{JP03}, I will consider the GSL to state that the generalized entropy is nondecreasing for all ``future causal horizons''.  A future causal horizon $H_{fut}$ is defined as $\partial \mathcal{I}^-(W_{fut})$, i.e. the boundary of the past of any future-infinite timelike worldline $W_{fut}$.  If this horizon is cut by some complete time slice $T$, the generalized entropy is the quantity:
\begin{equation}\label{gen}
S = \frac{A}{4G\hbar} + S_{out},
\end{equation}
where Boltzmann's constant $k$ is set to unity, $A$ is the expectation value\footnote{Taking the expectation value of the area is necessary to make the generalized entropy a c-number.  See Ref. \cite{10proofs} for discussion of this point.} of the area of $H_{fut}\,\cap\,T$, and $S_{out}$ is the fine-grained entropy stored in $\mathcal{I}^-(W_{fut})\,\cap\,T$, the region outside the horizon.  The only form of coarse graining used to define $S_{out}$ is the restriction of the system to the region outside the horizon.  (A naive calculation of $S_{out}$ is ill-defined due to divergences in the entanglement entropy near the horizon.  This divergence will be ignored until section \ref{ren}, in which I argue that the renormalization of $S_{out}$ should not affect the validity of the proof given in section \ref{proof}.)  Because of the ``many-fingered'' nature of time in general relativity, there are many different ways to push a time slice $T_1$ forwards in time to a new complete slice $T_2$ which is nowhere to the past of $T_1$.  The GSL states that there must be at least as much generalized entropy at $T_2$ as there was at $T_1$.

I will also need to use the time-reversal of the GSL.  This anti-GSL states that the generalized entropy of a past causal horizon $H_{past} \equiv \partial \mathcal{I}^+(W_{past})$ is non-\emph{increasing} with time, where $W_{past}$ is a past-infinite timelike worldline.  The anti-GSL can be deduced from the GSL together with CPT symmetry (the GSL is only affected by the time-reversal operation $T$ because it makes no reference to either spatial orientation or particle charges).  However, there is no need to assume CPT for the result below, as long as both the GSL and anti-GSL hold.

Assuming the anti-GSL may seem very strange.  Isn't the entire point of the second law that it favors one direction of time over the other?  This objection presumes that the generalized second law holds for the same reason as the ordinary second law.  For the ordinary second law, entropy increases with time as the result of the time-asymmetric assumption of low entropy initial conditions, together with some nontrivial coarse-graining procedure.  But for the GSL the time-asymmetric condition is simply that it holds on future horizons, whereas the anti-GSL holds on past horizons.  Furthermore, in order for the entropy to increase it is not necessary to use any other coarse-graining procedure besides restricting to outside of the horizon \cite{10proofs}.  Because of these two differences from the ordinary second law, it is reasonable to hope that the GSL can be proven without needing special time-asymmetric restrictions on the state.  (Note that in the absence of any past or future horizons, the fine-grained entropy is \emph{constant} with time, so both the GSL and the anti-GSL are satisfied.

In currently accepted cosmology there appear to be no true past horizons, since all worldlines are finite to the past because of the Big Bang singularity.  Nor are there any white holes, but there are approximate past Rindler-like horizons cutting through isolated astrophysical objects such as stars whenever the surrounding spacetime can be treated as asymptotically flat.

\section{The Structure of the Classical Background}\label{class}

By a ``classical background'', I mean a Lorentzian manifold $M$ which satisfies the Einstein equation
\begin{equation}\label{Ein}
 8\pi G\,T_{ab} = G_{ab}
\end{equation}
with respect to the stress energy tensor $T_{ab}$ of any classical matter fields which may happen to be on $M$.  These classical matter fields are required to obey the null energy condition
\begin{equation}
T_{ab}\,k^a k^b \ge 0
\end{equation}
for every null vector $k^a$ on $M$.  This is a natural restriction because classical theories which satisfy the Einstein equation but violate the null energy condition typically also violate the GSL, because of the possibility of sending negative energy fluxes into black holes.

Furthermore, there must be at least one null line $N$ contained in $M$ (or else the statement that the ANEC holds on all null lines is trivially true).  By the ``null splitting theorem'' of Galloway \cite{galloway00}, the existence of even a single null line $N$ in such a spacetime implies that $N$ is contained in a smooth, closed, achronal horizon $H \subseteq \partial \mathcal{I}^-(N)\,\cap\,\partial \mathcal{I}^+(N)$ which is unchanging with time (i.e., the zeroth order expansion parameter $\theta$ and shear tensor $\sigma_{ab}$ both vanish everywhere on $H$).\footnote{This theorem also requires a version of cosmic censorship: namely that every future causal horizon is null-geodesically complete to the future, and every past causal horizon is null-geodesically complete to the past.  This would follow from the ``slightly weaker extrastrong'' version of cosmic censorship found in Penrose \cite{penrose99}.}  Since there exist accelerating timelike worldlines $W_{fut}$ and $W_{past}$ which get closer and closer to $N$ in either the asymptotic future or the asymptotic past, $H$ is a connected component of both a past and a future causal horizon.

Because $H$ is achronal, the remainder of the spacetime $M - H$ is the union of three disjoint regions: the open region $P \equiv \mathcal{I}^-(H)$ to the chronological past of $H$, the open region $F \equiv \mathcal{I}^+(H)$ to the chronological future of $H$, and possibly a third ``other'' closed region $O \equiv M - (P \cup F \cup H)$ which is spacelike separated from all points on $H$.

One might worry that because Galloway's result is so strict, the implication is that very few spacetimes have null lines and thus the proof of the ANEC here applies only to special states, e.g. Kerr-Newman-(Anti-)de Sitter spacetime or pp-wave spacetimes.  However, it is important to note that these restrictions only apply to the classical background spacetime.  There is no requirement that the gravitationally perturbed spacetime contain any null lines or possess any symmetries.  Thus there is a large class of generic quantum states which are required by this proof to satisfy the ANEC.  This extends the proof of Wald and Yurtsever \cite{WY91} to arbitrary interacting fields (using the extra assumption that the GSL holds).

On the classical background metric, both the ANEC and the GSL inequalities are saturated.  Using the fact that $\theta$ and $\sigma_{ab}$ vanish, the Raychaudhuri equation
\begin{equation}\label{Ray}
-\frac{d\theta}{d\lambda} = \frac{1}{2}\theta^2 
+ \sigma_{ab}\sigma^{ab}
+ 8\pi G\,T_{ab} k^a k^b
\end{equation}
shows that $T_{ab} k^a k^b = 0$ on $H$.  Furthermore, if one assumes that any flux of outside entropy due to the classical matter fields must be supported by at least a little energy flux, then no matter entropy can cross the horizon either.  Furthermore, since $\theta = (1/A)(dA/{d\lambda})$, the area of the horizon is constant.  So the generalized entropy on the background is a constant classically; i.e., its time derivative has no component of order $\hbar^{-1}$.  Because the GSL is an inequality, this exact saturation on the classical background is needed for the GSL to be a nontrivial constraint at the next order of perturbation theory, which takes the effects of the quantum fields into account.

\section{The Perturbation Expansion}\label{pert}

This section describes the perturbation of the classical background $M$ due to quantum fields residing on it, which are in some particular state $\rho$.  A quantum field with order unity quanta excited has a stress-energy tensor of order $\hbar \lambda^{-4}$, where $\lambda$ is the wavelength of the excitations.\footnote{This statement depends on the choice of a reference frame, in which the size of the stress-energy tensor and the wavelength of the fields can be evaluated, since a large Lorentz boost can turn small quantities into large ones and vice versa.  Given a choice of unit timelike vector field $u^a$ at every point of spacetime, one can require that the metric and all other fields be slowly varying with respect to both the derivatives parallel to $u^a$ and perpendicular to $u^a$.  The order of magnitude of a tensor quantity can then be defined choosing the components of each index to be parallel or perpendicular to $u^a$.  The frame-dependent statements should then be read as though they were prefaced by: ``There exists at least one choice of $u^a$ such that...''}
Its gravitational effect on the curvature is of order $G\hbar \lambda^{-4}$.  Since the curvature is a second derivative of the metric, the effects on the metric are of order $G\hbar \lambda^{-2} = {l_P^2}/{\lambda^2}$, a dimensionless quantity.  Situations in which this is much less than one can be accurately described using QFT in curved spacetime, plus small corrections in powers of ${l_P^2}/{\lambda^2}$.  I will now choose units in which $G = 1$ and $\lambda \sim 1$, so that this expansion can be regarded as an expansion in Planck's constant $\hbar$.

The metric and the stress energy tensor may be expanded out as follows: 
\begin{align}\label{g}
g_{ab} &= g_{ab}^{0} + g_{ab}^{1/2} + g_{ab}^{1} + \mathcal{O}(\hbar^{3/2}), \\ \label{T}
T_{ab} &= T_{ab}^{0} + T_{ab}^{1}(\rho) + \mathcal{O}(\hbar^{3/2}).
\end{align}
Here and below, superscript numbers in parentheses indicate the number of powers of $\hbar$ contained in any quantity.  The Einstein equation relates the stress-energy tensor to the metric as follows:
\begin{align}\label{Ein0}
8\pi T_{ab}^{0}& = G_{ab}^{0}(g_{ab}^{0}), 
\\ \label{Ein1/2}
0& = \langle G_{ab}^{1/2} (g_{ab}^{0},\,g_{ab}^{1/2}) \rangle,
\\ \label{Ein1}
8\pi \langle T_{ab}^{1} \rangle & = \langle G_{ab}^{1} 
(g_{ab}^{0},\,g_{ab}^{1/2},\,g_{ab}^{1}) \rangle,
\end{align}
where $G_{ab}^{n}$ is defined as the Einstein tensor calculated from the metric using the usual formula, but keeping only terms which are of the exact order $\hbar^{n}$.  $G_{ab}^{n}$ is therefore a function only of the terms in the metric up to and including $g_{ab}^{n}$.\footnote{Notice that the expectation value has been taken of the first-order correction to the metric as well as the stress-energy tensor.  This means that I do \emph{not} need to use the semiclassical Einstein equation $G_{ab} = 8\pi \langle T_{ab} \rangle$, but only take the expectation value of both sides of the regular Einstein equation $\langle G_{ab} \rangle = 8\pi \langle T_{ab} \rangle$.  This can be done because the GSL as formulated in section \ref{gsls} refers only to the expectation value of the area.  Another consequence is that the results are not restricted to the case in which the fluctuations of the stress-energy tensor are small compared to the expectation value of the stress-energy tensor.}

The different order terms in the metric can be explained physically as follows:
\paragraph{Zeroth Order}
The zeroth order metric $g_{ab}^{0}$ is the finite (i.e. zeroth order in $\hbar$) background metric, which may be sourced by the stress-energy $T_{ab}^{0}$ of classical fields satisfying the null energy condition, as described in section \ref{class}, subject to the constraint that $T_{ab}^{0} k^a k^b = 0$ on $H$.

\paragraph{Half Order}
The order $\hbar^{1/2}$ term in the metric describes the metric fluctuations due to gravitons.  This is because the total energy of a gravitational wave scales as the metric perturbation squared, which is equal to $\hbar / \lambda$ for an order unity number of gravitons.  (Since gravitons do not contribute to the \emph{local} stress-energy tensor, there is no corresponding $\hbar^{1/2}$ term in $T_{ab}$.)  There is some gauge freedom available in choosing the metric perturbation term $g_{ab}^{1/2}$ due to diffeomorphism symmetry.

\paragraph{First Order}
The first order term in the stress-energy is due to the introduction of the quantum fields, which depends on the state $\rho$ of these fields.  This causes a small gravitational perturbation $g_{ab}^{1}$ which is also of order $\hbar$.  Given a particular stress-energy profile, there are multiple first order metric perturbations $g_{ab}^{1}$ which are consistent with Eq. (\ref{Ein1}).  This is partly due to the fact that general relativity admits gravitational wave solutions, and partly due to the diffeomorphism freedom of the perturbed metric, since specifying $g_{ab}^{1}$ requires a somewhat arbitrary identification of the spacetime points of the perturbed and unperturbed spacetimes.  However, any choice of $g_{ab}^{1}$ is permissible for purposes of the proof of the ANEC, so long as it is first order in $\hbar$ (if it is of larger magnitude than this, it must be included in the background metric $g_{ab}^{0}$).  The reason the ambiguity does not matter is that the only dependence of the generalized entropy on $g_{ab}^{1}$ is the expectation value of the first order correction to the area $A^{1}$, which will be related to the stress-energy tensor by means of the Raychaudhuri Eq. (\ref{Ray}).

\paragraph{Higher Orders}
Terms of $\hbar^{3/2}$ order and higher will be neglected.  These terms exist for two reasons:  First, the nonlinearity of the Einstein tensor in the metric makes terms nonlinear in the quantum stress-energy appear, which are of order $\hbar^2$ or higher.  Secondly, the small perturbations to the metric at half and first order affect the dynamics of the quantum fields, modifying their stress-energy tensor.  

\paragraph{}
In order to write down the GSL and anti-GSL on the spacetime manifold, it is necessary to determine the location of the past and future horizons.  Assume that the horizon is stable in the sense that it is not removed from the spacetime entirely in the perturbed spacetime.\footnote{This is usually true, but in some cases the global horizon may be removed by even a slight perturbation.  For example, if $M$ contains compact extra dimensions, even a small perturbation can sometimes result in some extra dimensions collapsing to a singularity after a finite though arbitrarily long time \cite{penrose03}, thus negating any horizons.  In such cases, the proof of the ANEC given below does not apply.  It might still be possible to prove the ANEC in such cases if one could place a bound on the maximum GSL violation possible for causal surfaces which are in some sense ``very close'' to being horizons.}  In order to be able to identify the past and future horizons on the perturbed manifold, it is also necessary to assume that the flux of null energy $T_{ab} k^a k^b$ and gravitational wave energy $\sigma_{ab}\sigma^{ab}$ falling across the horizon falls off sufficiently quickly to the asymptotic past and future of $H$ that the horizon may be taken as stationary in the asymptotic past and future.  That is, there must exist at least one identification of points between the perturbed and background spacetimes, such that at zeroth order all the fields are identical, while up to first order in $\hbar$,
\begin{equation}\label{bound}
\theta |_{H^+} = 0,\quad \theta |_{H^-} = 0
\end{equation}
where $H^+$ is the asymptotic future of $H$, and $H^-$ the asymptotic past of $H$.  (If there is more than one identification satisfying this property any of them may be selected in order to prove the ANEC.)  This identification can be used to define the future horizon $H_{fut}$ as that connected component of $\partial \mathcal{I}^-(H^+)$ which coincides with $H$ at zeroth order, and $H_{past}$ as that connected component of $\partial \mathcal{I}^+(H^-)$ which coincides with $H$ to zeroth order.  At zeroth order $H_{fut} = H_{past}$, but at first order $H_{fut}$ and $H_{past}$ usually separate into two distinct surfaces, one of which obeys the GSL while the other obeys the anti-GSL.

Do $H_{fut}$ and $H_{past}$ also separate at half order, or can they be taken to coincide until first order effects are considered?  This question will become important when gravitational fluctuations are considered in section \ref{graviton}, in which I will argue that the horizons cannot separate for nonextremal black holes or pp-wave spacetimes, and will prove a generalization of the ANEC under the assumption that they do not separate.

\section{Proof of the ANEC}\label{proof}

Having finished laying out the framework, I will now proceed to the core result of this paper: a proof of the ANEC for the null line $N$.  I will begin by ignoring the order $\hbar^{1/2}$ graviton fluctuations and the need to renormalize the generalized entropy, though neither of these can really be ignored in general.  I will discuss how to remedy these issues in the sections \ref{ren} and \ref{graviton}.

\begin{figure}[ht]
\centering
\includegraphics[width=\textwidth]{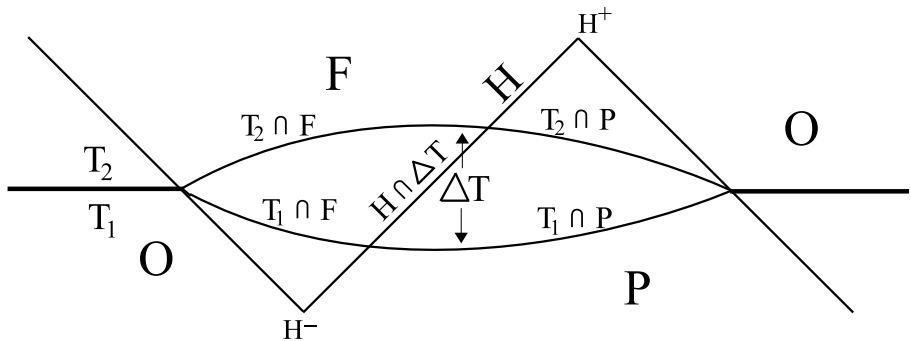}
\caption{\small{A Penrose diagram of the unperturbed classical spacetime.  $H$ is the horizon, shown here as a segment reaching from a past timelike infinite point $H^-$ to a future timelike infinite point $H^+$.  $P = \mathcal{I}^-(H)$ is the open region to the chronological past of $H$, and $F = \mathcal{I}^+(H)$ is the open region to the chronological future of $H$.  There may also be a closed region $O$ which is spacelike to $H$.  The slices $T_1$ and $T_2$, which coincide in $O$, are also shown.  $\Delta T$ is the spacetime region between the two slices.  Some specific intersections of regions important to the proof are also shown.  The boundary of the spacetime is not shown, except for the past and future of the horizon $H$.}}\label{causal}
\end{figure}

Let $T_1$ represent a complete time slice of $M$, $T_2$ represent another complete time slice nowhere to the past of $T_1$, and let $\Delta T$ represent the spacetime region between the two slices.  Let $T_2$ coincide with $T_1$ in the $O$ region so that during the interval $\Delta T$, any information which exits $P$ must flow into $F$ rather than into $O$ (see Figure \ref{causal}).

The $\hbar^0$ order contribution to the generalized entropy of the future horizon at a time slice $T$ depends on the zeroth order outside entropy and the first order contribution to the expected area:
\begin{equation}\label{S1}
S_{fut}^{0} = \left[ S_{out}^{0}[\,\rho\,](P\,\cap\,T) 
+ \frac{A^{1}}{4\hbar}[\,g_{ab}^{0},\,g_{ab}^{1}\,](H_{fut}\cap\,T) \right].
\end{equation}
Here square brackets represent the dependence of an extensive quantity on \emph{fields}, while round parentheses represent the dependence of the extensive quantity on \emph{regions} of spacetime.\footnote{There is a benign ambiguity in Eq. (\ref{S1}) due to the fact that in the first term the time slice $T$ is defined on the unperturbed spacetime, while in the second term $T$ must be a time slice on the perturbed spacetime.  This implicitly requires the identification of points in the perturbed and unperturbed manifolds, which as discussed in section \ref{pert} involves an arbitrary gauge dependency.  Fortunately the dependence of $A^{1}$ on this arbitrary choice is of order $\hbar^2$ and is therefore negligible.}

Since the $\hbar^{-1}$ order contribution to the generalized entropy is unchanging (cf. section \ref{class}), the GSL requires that the next highest order $\hbar^{0}$ contribution be nondecreasing:
\begin{equation}\label{+GSL}
\Delta \left[ S_{out}^{0}(P\,\cap\,T) + \frac{A^{1}}{4\hbar}(H_{fut}\cap\,T) \right] \ge 0,
\end{equation} 
where $\Delta f(T) \equiv f(T_2) - f(T_1)$.  By the time reverse of the above, the anti-GSL requires that
\begin{equation}\label{-GSL}
\Delta \left[ S_{out}^{0}(F\,\cap\,T) + 
\frac{A^{1}}{4\hbar}(H_{past}\cap\,T) \right] \le 0.
\end{equation}
Subtracting Eq. (\ref{-GSL}) from Eq. (\ref{+GSL}),
\begin{align}\label{dif1}
& \Delta \left[ S_{out}^{0}(P\,\cap\,T) 
- S_{out}^{0}(F\,\cap\,T) \right] \\ \label{dif2}
+\,\Delta &\left[ \frac{A^{1}}{4\hbar}(H_{fut}\cap\,T)
- \frac{A^{1}}{4\hbar}(H_{past}\cap\,T) \right] \ge 0.
\end{align}

The next step is to invoke the property of ``weak monotonicity'', which states that for any three regions $A$, $B$, and $C$ which are disjoint subsystems (i.e. they are mutually achronal and non-coincident), the entropy evaluated inside these regions must satisfy
\begin{equation}\label{weak}
S_{A\,\cup\,C} + S_{B\,\cup\,C} \ge S_{A} + S_{B}.
\end{equation}
Intuitively speaking, this property expresses that if the quantum system $C$ is more strongly entangled with $A$ than is allowed for by classical mechanics, this entanglement is ``private'', limiting the amount by which $C$ can be entangled with a third system $B$ (and in the extreme case where $C$ and $A$ are maximally entangled systems, $B$ cannot be entangled with $C$ at all).  Weak monotonicity is equivalent to the strong subadditivity property \cite{pipp03}, and is therefore true for general QFT states, modulo caveats about the divergence of the entanglement entropy, which will be addressed in section \ref{ren}.\footnote{Weak monotonicity should still apply even when, as below, the region $C$ is a null hypersurface.  This can be shown by proving strong subadditivity via the monotonicity of the mutual information, which holds quite generally due to its relation with the relative entropy \cite{CH04}.} In the present instance, if one takes the three regions to be:
\begin{eqnarray}
A &=& T_1\,\cap\,F, \\
B &=& T_2\,\cap\,P, \\
C &=& H\,\cap\,\Delta T,
\end{eqnarray} 
and uses the unitary property of QFT, then Eq. (\ref{weak}) becomes
\begin{equation}\label{weakapp}
S_{out}(T_2\,\cap\,F) + S_{out}(T_1\,\cap \,P) 
\ge S_{out}(T_1\,\cap\,F) + S_{out}(T_2\,\cap \,P).
\end{equation}
By collecting the $F$ and $P$ terms together, one obtains
\begin{equation}\label{nonpos}
\Delta \left[ S_{out}^{0}(P\,\cap\,T) 
- S_{out}^{0}(F\,\cap\,T) \right] \le 0,
\end{equation}
which shows that line (\ref{dif1}) is nonpositive.  Thus line (\ref{dif2}) is nonnegative:
\begin{equation}\label{deltaA}
\Delta \left[ A^{1}(H_{fut}\cap\,T) - A^{1}(H_{past}\cap\,T) \right] \ge 0.
\end{equation}

Eq. (\ref{deltaA}) says that to first order, the expected area of the future horizon always increases more over a time interval $\Delta T$ than the expected area of the past horizon does.  But this can only be true for all intervals $\Delta T$ if the expansion parameters satisfy
\begin{equation}\label{thetas} 
\langle \theta^{1}_{fut} - \theta^{1}_{past} \rangle \ge 0
\end{equation}
everywhere.  Now let $X$ be any point on the original null line $N$.  With the help of the linearized Raychaudhuri equation:
\begin{equation}\label{linray}
\frac{d \langle \theta^{1} \rangle }{d\lambda} = -8\pi \langle T_{ab}^{1} \rangle k^a k^b,
\end{equation}
and the boundary conditions from Eq. (\ref{bound}) appropriate for matching the perturbed horizon to the background horizon far from the perturbation:
\begin{equation}
\langle \theta^{1}_{fut} \rangle |_{\lambda = \infty} = 0,\quad \langle \theta^{1}_{past} \rangle |_{\lambda = -\infty} = 0,
\end{equation}
one may solve for $\theta$ in Eq. (\ref{thetas}):
\begin{equation}\label{ANEC2}
\int_X^\infty \langle T_{ab}^{1} \rangle \, k^a k^b\,d\lambda 
\,+\, \int_{-\infty}^X \langle T_{ab}^{1} \rangle \,k^a k^b\,d\lambda \,\ge\, 0,
\end{equation}
thus proving that the ANEC holds to first order in $\hbar$.  This is the same as saying that the ANEC holds exactly in the context of QFT on the background spacetime.  As a corollary, the ANEC integral can be zero only if the GSL (\ref{+GSL}), anti-GSL (\ref{-GSL}), and weak monotonicity (\ref{weak}) inequalities are all saturated.  Note also that to prove the ANEC, it is sufficient if the GSL and anti-GSL hold at even a single point $X$.\footnote{More precisely, the GSL and anti-GSL must hold for variations of $T$ which are confined to some neighborhood of $X$.}

\section{Renormalization}\label{ren}

In the last section the quantity $S_{out}$ was treated as a finite quantity despite the fact that in QFT, the naive von Neumann entropy diverges due to the entanglement of ultraviolet quantum field degrees of freedom near the boundary of the region.  For the GSL to have physical meaning, some way of making $S_{out}$ finite in the generalized second law will have to be found.  In the semiclassical limit, this could be done in a two step process similar to the way loop divergences are usually renormalized.  First one would have to regulate the entropy by imposing some sort of cutoff parametrized by a distance scale $\delta$, leading to a regulated entropy $S_\delta$.  The next step would be to subtract off the divergent part of $S_\delta$ by absorbing it into the coupling constants of the theory (in this case, various corrections to the Bekenstein-Hawking formula $A/{4\hbar}$ for the horizon entropy).  Finally, one would take the $\delta \to 0$ limit of the remaining convergent part of $S_\delta$, obtaining a finite, renormalized value $\tilde{S}_{out}$.  Some specific calculations of the renormalization of the generalized entropy have already been made \cite{FS94}, but I will not choose a specific approach here.  Instead I want to argue that the proof in section \ref{proof} will continue to hold for any renormalization scheme with certain reasonable properties, when the renormalized $\tilde{S}_{out}$ is used in place of the naive entropy $S_{out}$.

The first property needed is that the GSL should hold for the generalized entropy $S = A/{4\hbar} + \tilde{S}_{out}$.  Since the GSL is not even properly defined without a renormalization scheme, there has to be at least one renormalization scheme for which this is true or the GSL itself would be false or meaningless.  Since I am assuming in this proof the validity of the GSL, the substitution of $\tilde{S}_{out}$ into Eq. (\ref{+GSL}) and (\ref{-GSL}) is valid.

The only remaining question is whether the renormalized entropy satisfies the weak monotonicity property, which is needed to derive Eq. (\ref{weakapp}).  Since weak monotonicity is a general feature of every quantum mechanical system \cite{pipp03}, as long as the cutoff replaces the QFT with another unitary quantum mechanical theory (and as long as the observables in causally separated regions continue to commute) the cut-off regulated entropy $S_\delta$ will continue to satisfy weak monotonicity.

Finally, since obtaining the renormalized entropy $\tilde{S}_{out}$ from $S_\delta$ requires subtracting off the divergent part of $S_\delta$, it must be shown that this does not invalidate weak monotonicity.  To do this I will assume that the divergent terms subtracted off are: (i) the sum of terms which are associated with connected components of the boundary of the region, and (ii) the same on both sides of the connected component of the boundary.  Assumption (i) is motivated by the idea that the divergence of the entanglement entropy is entirely due to the UV degrees of freedom near the boundary, and therefore cannot depend on the the existence of another boundary a finite distance away.  Assumption (ii) is motivated by the fact that entropy satisfies the triangle inequality $S_{A\,\cup\,B} \ge |S_A - S_B|$.  If $B$ is the complement of $A$, $S_{A\,\cup\,B}$ should be finite, which implies that the divergent parts of $S_A$ and $S_B$ must be equal.  Given assumptions (i) and (ii), the divergent parts of $S_{out}$ cancel out in Eq. (\ref{weakapp}), because the boundaries $T_1\,\cap\,H$, $\,T_2\,\cap\,H$, and $T_1\,\cap\,\partial O\,(= T_2\,\cap\,\partial O)$ each appear twice, once positively and once negatively.

Therefore, the proof of the ANEC remains just as valid when renormalization is taken into account, so long as the renormalization scheme obeys the well-motivated axioms above.  The assumption most likely to be troublesome is the existence of a unitary regulated theory, since there are many common regulators that are not unitary (e.g. the Pauli-Villars regulator used by Demers, Lafrance, and Myers \cite{FS94}).   However, the other requirements are not very stringent (in particular there is no need for the regulator to preserve local Lorentz symmetry), so it seems likely that an acceptable regulator exists.

\section{Gravitational Fluctuations}\label{graviton}

Can the ANEC be generalized to include gravitational null energy?  Since no local stress-energy tensor can be assigned to gravity waves, in general there is no well defined integrated null energy.  However, in the special case of spacetimes possessing a null line, one can define the gravitational null energy falling across a horizon as the shear squared $\sigma_{ab}\sigma^{ab}$ of the past or future horizons $H_{past}$ or $H_{fut}$, which contributes to the Raychaudhuri equation in a similar way to the stress-energy tensor of matter.  Although naively this term is positive, the infinite subtraction needed to make it finite permits situations in which $\sigma_{ab}\sigma^{ab} < 0$ \cite{CS77}.  Since this has similar effects to a violation of the null energy condition, in order to rule out traversable wormholes, closed timelike curves, and negative energy states, it is necessary to show some analogue of the ANEC for gravitational energy.

The first step is to show that at $\hbar^{1/2}$ order, gravitational wave perturbations cause no separation between the past and future horizons $H_{past}$ and $H_{fut}$.  I suspect that this is always true, but I can only prove it in the cases when the null line $N$ lies on the horizon of a nonextremal asymptotically flat black hole, or in a pp-wave spacetime.

Since at this order the graviton is a free field, the linear field operators satisfy the classical field equation, so to prove that the horizons do not separate it will be sufficient to consider classical gravity wave solutions, which must obey the null energy condition (because they do not contribute to the stress-energy tensor at all).  Two different cases will be briefly sketched:
\begin{enumerate}
\item \label{i} If $N$ lies on the horizon of a nonextremal asymptotically flat black hole, the horizon generators of $H$ are marginally trapped null lines threading a wormhole in every such black hole solution of which I am aware (for vacuum solutions the Kerr case is the most general).  $H^-$ therefore lies on the boundary of $\mathcal{I}^-$ on one side of the wormhole, and $H^+$ lies on the boundary of $\mathcal{I}^+$ on the other side of the wormhole.\footnote{If as the result of nontrivial topology the $\mathcal{I}^+$ and $\mathcal{I}^-$ lie in the same asymptotic region, as in the case of the $RP^3$ geon, the horizon generators will no longer be achronal.}  Since the half order fluctuations obey a linear equation, if any separation of horizons is possible then by choosing the sign of the perturbation one could arrange that $H_{fut}$ lies inside $H_{past}$ somewhere.  But since $H_{fut}$ lies on the boundary of $\mathcal{I}^+$ and $H_{past}$ lies on the boundary of the $\mathcal{I}^-$ on the other side of the wormhole, this would make the $\mathcal{I}^+$ lie to the future of the $\mathcal{I}^-$, thus making the wormhole traversable.  So there can therefore be no separation of horizons at half order without violating the already established classical version of the topological censorship theorem in Ref. \cite{FSW93}, assuming that the half order $\hbar$ expansion is a good approximation to the full nonperturbative classical solution to which the theorem is applied.

\item If $N$ lies in a pp-wave spacetime, there are multiple possible choices of $H_{fut}$ and $H_{past}$, but the choice of one determines the choice of the other if one requires $H_{fut}$ and $H_{past}$ to coincide for horizon generators asymptotically far from the perturbation.  Then a similar argument to case (\ref{i}) may be used, except that in this case if $H_{fut}$ lies inside $H_{past}$, the no superluminal communication theorem of Ref. \cite{olum98} is violated instead.  (When the pp-wave spacetime is the Minkowski vacuum, the no superluminal communication result of Ref. \cite{VBL99} is also violated.)
\end{enumerate}
Because these two cases cover a diverse range of spacetimes with null lines, it is reasonable to conjecture that in general there can be no separation of $H_{past}$ and $H_{fut}$ at half order.

Assuming this conjecture, it is now possible to show a generalization of the ANEC valid when there are gravitational fluctuations.  Since $H_{past} = H_{fut}$, it follows that the shear tensor $\sigma_{ab}$ is also identical at half order on both horizons.  The first order Raychaudhuri equation states that
\begin{equation}\label{shearR}
-\frac{d \langle \theta^{1} \rangle}{d\lambda} = 
\langle \sigma_{ab}\sigma^{ab} \rangle^{1}
+ \langle R_{ab} \rangle^{1} k^a k^b,
\end{equation}
where the $\theta^2$ term is omitted since it is second order in $\hbar$.  I have not specified which renormalization scheme is used to regulate the divergences in Eq. (\ref{shearR}).  But since Eq. (\ref{shearR}) is a tautology when both sides are defined in terms of functions of the metric on the horizon and its derivatives, so long as the metric variables implicit in the terms of Eq. (\ref{shearR}) are renormalized in the same way in each term, and the same operator ordering prescription can be used, Eq. (\ref{shearR}) should still be true.  One can then use the Einstein equation to rewrite Eq. (\ref{shearR}) in terms of the stress-energy tensor:
\begin{equation}\label{shearray}
-\frac{d \langle \theta^{1} \rangle}{d\lambda} = 
\langle \sigma_{ab}\sigma^{ab} \rangle^{1}
+ 8\pi \langle T_{ab} \rangle^{1} k^a k^b.
\end{equation}

The proof in section \ref{proof} still works up to the derivation of Eq. (\ref{thetas}):
\begin{equation}
\langle \theta^{1}_{fut} - \theta^{1}_{past} \rangle \ge 0.
\end{equation}
But now $\theta$ is given by the integral along $N$ of Eq. (\ref{shearray}) instead of Eq. (\ref{linray}), and the resulting shear-inclusive version of the ANEC states that on the null line $N$:
\begin{equation}\label{anec3}
\int^{\infty}_{-\infty} (\langle T_{ab}^{1} \rangle k^a k^b + 
\frac{1}{8\pi} \langle \sigma_{ab}\sigma^{ab} \rangle^{1}) d\lambda \ge 0,
\end{equation}
where again the equality is only possible when the GSL, anti-GSL, and weak monotonicity integrals are all saturated.  Naively this equation appears to be weaker than Eq. (\ref{ANEC2}) but it is not, since the shear-squared can be negative.   Because of its dependence on $\sigma_{ab}$, Eq. (\ref{anec3}) is not even defined except on null lines, since otherwise the shear of the past and future horizons are not necessarily the same.  (Perhaps this is related to the fact that the ANEC can be violated on geodesics that are not null lines.)

In situations where the shear squared may be negative, it is the shear-inclusive ANEC rather than the regular ANEC which is better suited to proving theorems about topological censorship, positivity of energy, and the absence of closed timelike curves.  For spacetimes which violate any of these conditions must have null lines \cite{GO07}\cite{PSW93}, which by Eq. (\ref{anec3}) must have either a positive or zero shear-inclusive ANEC integral.  Since traversable wormholes, closed timelike curves, and negative total energies must by continuity be possible in generic spacetimes if they are possible at all, the integral can be taken to be positive rather than zero.

But if the left-hand side of Eq. (\ref{anec3}) is positive along any null geodesic $N$ lying on a null congruence, then $N$ is required to eventually either be singular or else have conjugate points.\footnote{These conjugate points would occur at very large values of the affine parameter, and would represent a breakdown of the applicability of Eq. (\ref{shearray}).  However, since the local metric perturbation is still assumed to be small even once the expansion parameter $\theta$ becomes large, the classical background Raychaudhuri Eq. (\ref{Ray}) can take over at that point, to show the existence of the conjugate point.}  In that case it cannot be a remain a null line on the perturbed spacetime, contradicting the initial supposition.  Therefore traversable wormholes etc. are ruled out by the shear-inclusive ANEC in perturbative quantum gravity whenever effects of order $\hbar^{3/2}$ and higher are negligible.

Despite the fact that the regular ANEC is not as conceptually important as the shear-inclusive ANEC when there are gravitational fluctuations, the former can also be proven from Eq. (\ref{anec3}) on the assumption that the GSL holds no matter how many species of particles there are.  Copy the matter fields $N_s$ times and put them all in the same state $\rho$.  (This is possible if the matter and gravity sectors are all noninteracting.)  Then Eq. (\ref{anec3}) implies that 
\begin{equation}
\int^{\infty}_{-\infty} (N_s \langle T_{ab}^{1} \rangle k^a k^b + 
\frac{1}{8\pi} \langle \sigma_{ab}\sigma^{ab} \rangle^{1}) d\lambda \ge 0.
\end{equation}
where $T_{ab}^{1}$ is the stress-energy of a single matter sector.  For $N_s = \hbar^{p}$, with $-1/2 < p < 0$, there are enough sectors to neglect the shear-squared term but not so many that the gravitational interactions between sectors become large.  Thus each individual matter sector satisfies Eq. (\ref{anec}), the regular ANEC.  Similarly, if $N_s = 0$ the only thing left is the shear-squared term, and a purely gravitational form of the ANEC is obtained.

\section{Conclusion}\label{con}

Any null line $N$ on a classical background lies on both a past and a future horizon.  Assuming that these horizons are perturbed slightly by quantum fields, the assumption that the GSL and its time-reverse are true near a single point on $N$ is sufficient to derive the ANEC on $N$ as well.  This conclusion depends on the existence of a suitable regulator for the generalized entropy, which has not been proven but was argued for in section \ref{ren}.  According to Graham and Olum \cite{GO07}, the achronal ANEC is sufficiently strong to obtain positivity of energy, the absence of closed timelike curves, and topological censorship, so that these conclusions also follow from the GSL under the conditions outlined in section \ref{intro}. 

However, these proofs are valid only under the assumption that $\sigma_{ab}\sigma^{ab} \ge 0$, which fails when quantum fluctuations of the metric are taken account.  To remedy this flaw one may include a shear-squared term in the achronal ANEC integral.  In section \ref{graviton}, this shear-inclusive ANEC was proven from the GSL for null lines lying on nonextremal black hole event horizons or in pp-wave spacetimes.  This result can be extended to all null lines if no gravity wave perturbation in classical general relativity can cause a past and future horizon to separate to linear order.  If the shear-inclusive ANEC integral is also generically positive, then the proofs of positivity of energy etc. are sound.

The situations in which we have a right to expect either form of the ANEC to hold are limited, since most background spacetimes do not possess null lines.  Since the GSL implies the ANEC in the situations where the ANEC is expected to hold, it might be illuminating to try to prove positivity of energies etc. directly from the GSL.  Previous proofs of the ANEC have mostly relied on the technical properties of particular field equations, but this work suggests that there is a more general conceptual reason for the ANEC.

It is a widely held view that the GSL must hold in any semiclassical approximation to a theory of quantum gravity, as a result of the underlying theory's statistical mechanics.  On this view the ANEC should hold for any set of minimally coupled matter fields capable of being given a UV-completion in a quantum gravity theory.  On the other hand, if there is a general semiclassical proof of the GSL, this would also prove the ANEC for all matter fields in QFT without reference to any UV-completion of gravity.  But such a proof would have to apply to the case of rapidly changing quantum fields, which has not yet been done \cite{10proofs}.

\small
\subsection*{Acknowledgments}

I would like to thank my advisor, Ted Jacobson, for many helpful conversations regarding both the form and the content of this paper, without whom the conclusions of this paper would have been significantly less developed.  I would also like to thank Bob Wald's General Relativity group for discussion at an early stage.  This work was also supported by the National Science Foundation grants PHY-0601800 and PHY-0903572, the Maryland Center for Fundamental Physics, and the Perimeter Institute.

\end{document}